   \documentclass[prl,aps,twocolumn,showpacs]{revtex4}
\usepackage[dvips]{graphicx}
\usepackage{dcolumn}%
\usepackage{amsmath}%
\usepackage{amsfonts}%
\usepackage{amssymb}
\providecommand{\U}[1]{\protect\rule{.1in}{.1in}}
\newcommand{\be}{\begin{equation}}
\newcommand{\ee}{\end{equation}}
\newcommand{\bea}{\begin{eqnarray}}
\newcommand{\eea}{\end{eqnarray}}
\newcommand{\bt} {\begin{tabular}}
\newcommand{\et} {\end{tabular}}
\newcommand{\nn}{ \nonumber}

\newcommand{\ba} {\begin{array}}
\newcommand{\ea} {\end{array}}
\begin{document}

\title{The effect of Coulomb  interactions on nonlinear thermovoltage and thermocurrent in quantum dots}

\author{Natalya A. Zimbovskaya}

\affiliation
{Department of Physics and Electronics, University of Puerto 
Rico-Humacao, CUH Station, Humacao, Puerto Rico 00791, USA; \\
 Institute for Functional Nanomaterials, University of Puerto Rico, San Juan, Puerto Ruco 00931, USA} 

\begin{abstract}
In the present work, we theoretically study the nonlinear regime of charge transport through a quantum dot coupled to the source and drain reservoirs. The investigation is carried out using a nonequilibrium Green's functions formalism beyond the Hartree-Fock approximation.  Employed approximations for the relevant Green's functions allow to trace a transition from Coulomb blockade regime to Kondo regime in the thermoelectric transport. Effects arising when electrons move in response to thermal gradient applied across the system are discussed, including experimentally observed thermovoltage zeros.
   \end{abstract}


\date{\today}
\maketitle

\section{i. introduction}

Thermoelectric properties of mesoscale and nanoscale systems are being intensively studies in the last two decades. Tailored nanostructures such as quantum dots (QD) and/or molecules sandwiched in between conducting electrodes attract significant research interest partly because they are expected to be useful in manufacturing of   highly efficient energy conversion devices. The improved efficiency of heat-to-electricity conversion in these systems originates from sharp features appearing in their electron transmission spectra, as it was predicted in earlier studies of thermoelectric properties of solids \cite{1,2,3}. Apart from possible applications, studies of thermoelectric properties of nanoscale systems can provide a deeper insight into the nature and characteristics  of electron and thermal transport mechanisms \cite{4,5,6,7}. Recently, a new research field of nanoscale thermoelectronics has emerged and thermoelectric properties of tailored  nanosystems have been explored both experimentally and theoretically \cite{8,9,10,11,12}.

As known, heat-to-electric-power converters operate due to Seebeck effect. The latter occurs when thermal and electric driving forces simultaneously affect electron transport through the considered system. A thermovoltage $ V_{th} $ appears when  the temperature differential $ \Delta \theta $ is applied across the unbiased system provided that the electric current is completely suppressed.  Thus the emergence of $ V_{th} $ indicates the energy conversion. Another quantity characterizing thermoelectric transport through QD, molecular  junctions and other nanoscale systems of similar kind is thermocurrent $I_{th}.$ It may be defined as a difference between the electron tunnel current flowing through a biased system in the presence of a temperature differential and the current flowing at $ \Delta \theta = 0. $ 
%% \be  I_{th} = I (V, T_0,\Delta T) - I(V, T_0, \Delta T = 0).      \label{1} \ee            Here, $T_0 $ is the original temperature of the system.
    As well as the thermovoltage, $I_{th} $ is controlled by simultaneously acting electric and thermal driving forces, and the combined effect of these forces depends on the bias voltage polarity and on the type of charge carriers involved in the transport process. Assuming for example that the left electrode of the considered junction is kept at higher temperature  than the right one, $ I_{th} $ takes on negative/positive values when charge carriers are correspondingly electrons/holes. %% Also, it was shown that (disregarding electron-phonon interactions) maximum efficiency of nanoscale heat-to electric energy converter could be reached when the electric current vanishes. Under this condition $ I_{th} = - I(V,T_0,\Delta T = 0). $ Therefore, the change of the sig of $ I_{th} $ indicates that the system turns from the regime favorable for energy conversion to that which is not favorable or vice verse.
   
   When $ \Delta \theta \ll \theta_{L,R}\ (\theta_{L,R} $  being the temperatures of the left and right electrode, respectively) the system operates within the linear in temperature regime, so $ V_{th} = -S\Delta T. $ Within this regime, the thermopower $ S $ describes the efficiency of energy conversion along with the thermoelectric figure of merit $ ZT. $ Correspondingly, properties of both $ S $ and $ ZT $ in nanoscale systems have been (and still are) intensively studied \cite{13,14,15,16,17,18,19,20,21,22}. As the temperature differential across the system increases, the system may switch to nonlinear regime of operation. For example, a thermovoltage that nonlinearly changes with $ \Delta \theta $ was observed in experiments on semiconductor QDs and single-molecule junctions \cite{23,24,25}. Presently, significant effort is being applied to theoretically analyze thermoelectric properties of nanoscale energy converters operating beyond linear regime \cite{26,27,28,29,30,31,32,33,34,35,36}.

It is known that Coulomb interactions between charge carriers may strongly affect thermoelectric properties of QD/molecules leading to novel and distinct phenomena. Manifestations of Coulomb interactions vary depending on relationship between the charging energy $ U, $ the coupling strength $ \tau $ characterizing  dot-leads contacts and the characteristic temperature $ \theta = \frac{1}{2}(\theta_L + \theta_R). $ When the dot/molecule is weakly coupled to the leads, so that $ U $ significantly exceeds both $ \tau $ and the thermal energy $ k\theta \ (k $ is Boltzmann's constant) the  equilibrium density of states of electrons on a single-level dot (DOS) displays two peaks whose separation equals $ U, $  as shown in the Fig. 1. As the coupling of the dot to the electrodes strengthens, the peaks in the DOS become lower and broader.  A furthermost growth of $ \tau $ results in disappearance of the Coulomb blockade peaks. When the temperature $ \theta $ takes on values below the Kondo temperature $ \theta_k ,$ a single sharp maximum emerges in their stead. At higher temperatures, the DOS becomes smooth and featureless. 

\begin{figure}[t] %%% fig. 1
\begin{center}
\includegraphics[width=8.8cm,height=4.5cm]{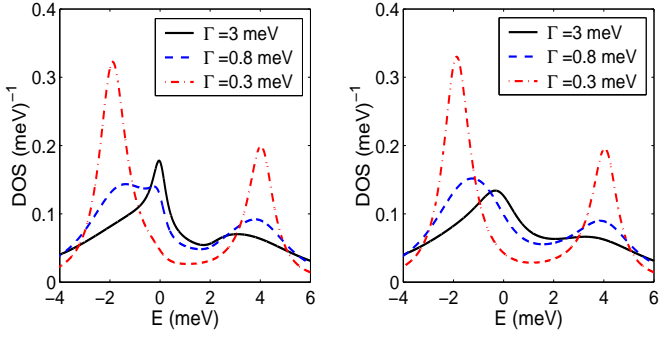} 
\caption{(Color online)  The equilibrium electron density of states  on a single-level quantum dot. The curves are plotted assuming that $E_0 = -2meV;\ U = 6meV;\ $ at $ k\theta_L = k\theta_R = 0.16 meV $ (left panel) and  $ k\theta_L = k\theta_R = 0.6 meV $ (right panel).
}
 \label{rateI}
\end{center}\end{figure}

The effects of Coulomb interactions on thermoelectric transport through QD/molecules were intensely studied (See e.g. Refs. \cite{15,16,20,26,27,37}). In the present work we contribute to these studies by analyzing the effect of Coulomb interactions on the nonlinear thermovoltage and thermocurrent. 

\section{I. Main Equations}

In the following analysis we concentrate  on electron contribution to thermoelectric transport omitting from consideration the phonon part. Retaining a single orbital on the dot/molecule, we write the relevant Hamiltonian as $ H = H_D + H_L + H_R + H_T. $ Here, the first term describes the dot. It is taken in the standard form:
\be
H_D = \sum_\sigma E_\sigma d_\sigma^\dag d_\sigma + Ud_\sigma^\dag d_\sigma d_{-\sigma}^\dag d_{-\sigma}.        \label{1}
\ee
Here, $d_\sigma^\dag,d_\sigma $ are creation and annihilation operators for the electrons on the dot with a certain spin orientation, $ E_\sigma = E_0 $ is the energy of a single spin-degenerated dot level and $ U $ is the charging energy. The terms $ H_\beta\ (\beta = L,R) $ are corresponding to noninteracting electrons on the left/right electrode:
\be
H_\beta = \sum_{r\sigma} \epsilon_{r\beta\sigma} c_{r\beta\sigma}^\dag c_{r\beta\sigma} \label{2}
\ee
where $\epsilon_{r\beta\sigma} $ are single-electron energies on the lead $ \beta $ and $ c_{r\beta\sigma}^\dag,\ c_{r\beta\sigma} $ are creation and annihilation operators for electrons on the leads. The last term:
\be
H_T = \sum_{r\beta\sigma} \tau_{r\beta\sigma} c_{r\beta\sigma}^\dag d_\sigma + H.C.               \label{3}
\ee
describes tunneling effects between the dot and the electrodes. The factors $ \tau_{r\beta\sigma} $ are coupling parameters characterizing the coupling of the electron states on the dot to the leads. For a symmetrically  coupled system $ \tau_{rL\sigma} = \tau_{rR\sigma} \equiv \tau_{r\sigma}. $

Now, we employ the EOM method to compute the retarded Green's function for the dot. Disregarding spin-flip processes, we arrive at separate expressions for the Green's functions corresponding to different spin channels. These expressions can be presented in the form:
\be
G_\sigma^{rr}(E) = \frac{E - E_0 - \Sigma_{02}^\sigma - U(1 - \left<n_{-\sigma}\right>)}{(E - E_0 - \Sigma_{0\sigma})(E - E_0 - U - \Sigma_{02}^\sigma) + U\Sigma_{1\sigma}} .   \label{4}
\ee
Here, $\left< n_{-\sigma}\right> $ are one-particle occupation numbers on the dot:
\be
\big<n_\sigma\big> = \frac{1}{2\pi} \int dE \mbox{Im} \big(G_\sigma^<(E)\big) .  \label{5}
\ee
Self-energy corrections included into the expression (\ref{4}) have the form:
\begin{align}  
 \Sigma_{0\sigma} =\, & \sum_{r\beta} \frac{|\tau_{r\beta\sigma}|^2}{E - \epsilon_{r\beta\sigma} + i\eta} \equiv \Sigma_{0\sigma}^L + \Sigma_{0\sigma}^R, \label{6}
\\ 
\Sigma_{1\sigma} =  & \sum_{r\beta} |\tau_{r\beta,-\sigma}|^2 f_{r,-\sigma}^\beta 
\bigg\{\frac{1}{E -  \epsilon_{r\beta,-\sigma} + i\eta}
\nn\\ & +
 \frac{1}{E - 2E_0 - U + \epsilon_{r\beta,-\sigma} + i\eta}   \bigg\},  \label{7}
\\ 
\Sigma_{2\sigma} = &  \sum_{r\beta} |\tau_{r\beta,-\sigma}|^2 
\bigg\{\frac{1}{E -  \epsilon_{r\beta,-\sigma} + i\eta}
\nn\\ & +
\frac{1}{E - 2E_0 - U + \epsilon_{r\beta,-\sigma} + i\eta}  \bigg\}, \label{8}
\\  
\Sigma_{02}^\sigma =  &\, \Sigma_{0\sigma} + \Sigma_{2\sigma} \label{9}
\end{align}
where $ f_{r\sigma}^\beta $ is the Fermi distribution function for the energy $\epsilon_{r\beta\sigma}$ and chemical potential $ \mu_\beta $ and $ \eta $ is a positive infinitesimal parameter. The expression (\ref{4}) was first obtained by Meir et al \cite{38}. Later, the same expression was derived and employed in several works where transport properties of quantum dots and molecules were theoretically studied. 

The lesser Green's function $G_\sigma^<(E) $ is related to the retarded and advanced Green's functions ($ G_\sigma^{rr} (E) $ and  $ G_\sigma^{aa} (E) ,$ respectively) by Keldysh equation:
\be
G_\sigma^<(E) = G_\sigma^{rr}(E) \Sigma_\sigma^< (E) G_\sigma^{aa} (E).
\label{10}  \ee
Both charge and energy transfer through a QD/molecule weakly coupled to the electrodes may be successfully studied employing NEGF within the Hartree-Fock approximation.This method brings simpler expressions for the Green's functions than those given by Eqs. (\ref{4})-(\ref{9}). Being employed to study current-voltage characteristics in a single-level quantum dot, it yields typical stair-like curves with two steps  maintaining the correct $\bf 2 : 1 $ ratio of heights of two successive steps for a dot symmetrically coupled to leads \cite{39}. However, the Hartree-Fock approximation cannot be  used to analyze charge/energy transport through a QD/molecule strongly coupled to the leads at low temperatures. To catch the Kondo peak and study related transport phenomena one needs to compute Green's functions beyond the Hartree-Fock approximation. To correctly describe transport within the Kondo regime, it is important to choose a suitable approximation for the lesser Green's functions. 

In further calculations, we approximate $ \Sigma_\sigma^<(E) $ as follows:
\be
\Sigma_\sigma^<(E) = i \sum_\beta f_\sigma^\beta(E) \Gamma_\sigma^\beta(E)              \label{11}
\ee    
where $ \Gamma_\sigma^\beta(E) = -2 \mbox{Im} [\Sigma^{0\beta}_\sigma(E)] $ and $ f_\sigma^\beta (E) $ is the Fermi distribution function for the left/right electrode. This approximation leads to the correct $\bf 2 : 1 $ ratio of heights of two steps displayed on the current-voltage curves for a symmetrically coupled system $ (\Gamma_\beta^L = \Gamma_\beta^R = \Gamma)$ within the limit of weak coupling $( \Gamma\ll U) ,$ as shown in an earlier work \cite{40}. On  these grounds, we believe that Eqs. (\ref{4})-(\ref{11}) may be employed to study characteristics of thermoelectric transport through QD/molecules in both Coulomb blockade (weakly coupled systems) and Kondo regime (strongly coupled systems below Kondo temperature). 

For a symmetrically coupled system including a single-energy QD/molecule as a linker, electric current is given by the Landauer expression:
\be
 I = \frac{e}{\pi\hbar} \int T(E) \big[f^L(E,\theta_L) - f^R(E,\theta_R)\big] dE.            \label{12}
\ee
Here, $ T(E) $ is the electron transmission function:
\be
 T(E) = \frac{i}{2}\Gamma \sum_\sigma \big[G_\sigma^{rr}(E) - G_\sigma^{aa}(E) \big].            \label{13}
\ee
where the wide-band approximation is used for the coupling parameter $\Gamma. $ The Green's functions included in (\ref{13}) implicitly depend on temperatures of the electrodes $ \theta_L $ and $ \theta_R $ because self-energy terms $ \Sigma_{1\sigma} $ given by Eq. (\ref{7}) include temperature dependent Fermi distribution functions $ f_{r\sigma}^\beta. $ Also, in the presence of temperature differential $ \Delta \theta = \theta_L - \theta_R $ the bridge energy $ E_0 $ may be renormalized including a correction proportional to $ \Delta \theta $ namely: $ \tilde E_0 = E_0 + z k \Delta \theta $ where $z $ is a dimensionless factor whose magnitude is of the order of one or smaller \cite{34}.  Therefore, the electron transmission depends of the temperatures $ \theta_L,\theta_R. $ In the following analysis we assume for certainty that the right electrode is kept at a fixed temperature, and it remains cooler than the left one, so $ \Delta\theta > 0. $ Basing on Eqs. (\ref{1})-(\ref{13}) one may theoretically analyze characteristics of thermoelectric transport through the considered  systems.

\section{iii. thermovoltage}

As known, the thermovoltage is determined from the open-circuit condition $ I(V_{th},\Delta\theta) = 0 $ assuming that no bias voltage is applied across the system. Solving this equation one obtains $V_{th}(\Delta\theta).$ In the present work $V_{th} $ was computed at two different values of $\theta_R.$ These values were chosen in such  a way that one of them is below the Kondo temperature whereas another one is above the latter. Obtained results are presented in the Fig. 2. At small values of temperature differential the thermovoltage is a linear function of $ \Delta\theta $ which can take on positive or negative values depending on the nature of charge-carriers (electrons/holes) involved in transport. As $ \Delta\theta $ increases, $V_{th} $ reveals a distinctly nonlinear  dependence of $ \Delta \theta. $ 

        With increasing $ \Delta \theta $ the thermovoltage magnitude increases, and it reaches its maximum/minimum at a certain value of $ \Delta\theta. $ Further increase of the temperature differential reduces  $ V_{th} $ magnitude. For every value of the bridge energy $ E_0, $ there exists  the corresponding value of $ \Delta\theta $ were $ V_{th} $ becomes zero. Subsequent growth of the temperature differential leads to emergence of thermovoltage with the reversed polarity.  %% All curves shown in the Fig. 2 are plotted assuming that the left electrode is hotter than the right one. Therefore, the curves presenting $ V_{th} $ taking on negative/positive values at small $ \Delta\theta $ correspond to the case when charge-carriers mainly contributing to thermally excited transport are electrons/holes.
				We remark that some curves plotted in the Fig. 2 qualitatively agree with those obtained in recent experiments on semiconducting QD \cite{25}.

The reversal of $ V_{th} $ polarity may be explained as follows \cite{40}. Assuming for certainty that chemical potentials of the electrodes in the unbiased system $ \mu_L = \mu_R = 0, $ and the charge carriers are electrons, the difference in the  temperatures of  electrodes leads to the flow of electrons from the left (hot) to the right (cool) electrode. To suppress this thermally excited current, a negative thermovoltage emerges which grows in magnitude as $ \Delta\theta $ increases. However, as $ \theta_L $ rises, the step in the Fermi distribution function for the left electrode is being partially smoothed out. This opens the way for holes to flow to the right electrode.  At certain value of temperature differential the hole flux completely counterbalances the electron flux. So, at this value of temperature differential the thermally excited electric current vanishes at $ V_{th} = 0. $

As known, the channels for thermally excited electron transport are opening up when a maximum in the DOS is located near the chemical potential of the electrodes. When we consider the system within the Kondo regime, the DOS reveals a single maximum near $ E = 0 $ which is shown in the left panel of the Fig. 1. So, the electron transport occurs through $ E =0 $ transport channel. However, both shape and position of the Kondo peak depend on the position of the bridge energy level $ E_0. $ Also, in the presence of a temperature  differential, $E_0 $  becomes renormalized itself. It appears that  even slight shifts in the bridge level position may result in significant variations of the intensity of electron transport through the channel associated with the Kondo peak. The more suitable for transport this channel becomes, the greater values $ V_{th} $ takes. Comparing the curves displayed in the left panel of Fig. 2, one observes that $ E_0 = - 1meV $ provides the most advantageous conditions for the thermally excited transport below the Kondo temperature than other values of $ E_0 $ used to plot these curves.

\begin{figure}[t] %%% fig. 2
\begin{center}
\includegraphics[width=8.8cm,height=4.5cm]{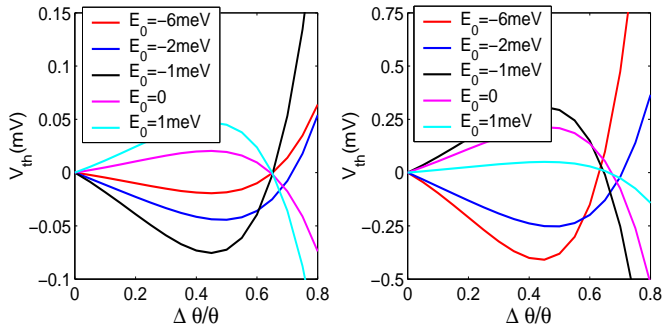} 
\caption{(Color online) Thermovoltage as a function of the temperature differential for a system below (left panel) and above (right panel) the Kondo temperature presented at several values of the bridge energy $ E_0. $  Left panel: The curves are plotted at $  U = 6meV,\ \Gamma = 3meV,\ k\theta_R = 0.16meV$  (left) and  $ \Gamma = 0.3meV,\ k\theta_R = 0.6meV $ (right).  
}
 \label{rateI}
\end{center}\end{figure}

When  the temperature of the coolest region rises above $ \theta_k, $ the Kondo peak in DOS disappears but two broader and more robust peaks corresponding to Coulomb blockade emerge provided that the coupling of the bridge to electrodes is sufficiently weak. In this case, there are two possible transport channels corresponding  to these peaks. Omitting from consideration the renormalization of the bridge energy $ E_0$  due to the temperature gradient one may conclude that the best conditions for thermally excited electron tunneling in an unbiased system occur when $  E_0 = -U $ and $  E_0 = 0. $ The curves presented in the right panel of the Fig. 2 confirm this conclusion. The greatest values of $ V_{th} $ are reached at $ E_0 $ close to $ - U $ and/or  zero. Slight discrepancies between the predicted above and computed values of of $ E_0 $ providing the best conditions for thermally excited transport originate from the fact that $ E_0 $ is renormalized by $ \Delta \theta $ and this was taken into account while computing the results shown in the Fig. 2.

\begin{figure}[t] %%% fig. 3
\begin{center}
\includegraphics[width=8.8cm,height=4.5cm]{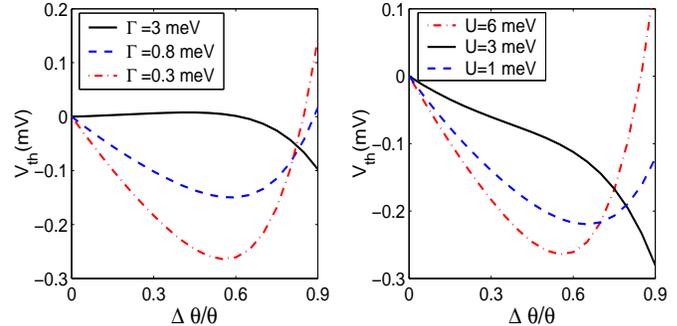} 
\caption{(Color online) Left panel: The effect of coupling of the electrodes to the bridge on the thermovoltage. The curves are plotted at $ U = 6meV,\ E_0 = - 1meV, k\theta_R = 0.6meV.$  Right panel: The thermovoltage as a function of temperature differential at several values of charging energy $U.$ Curves are plotted for $k\theta_R=0.6meV;\ E_0=-1meV;\ \Gamma = 0.3meV. $
}
 \label{rateI}
\end{center}\end{figure}

To further analyze the effect of Coulomb interactions between charge carriers on the bridge  we consider $ V_{th} (\Delta\theta) $ at several different values of $ \Gamma $ assuming that the system is kept at a relatively high temperature $(\theta_R > \theta_k).$ As presented in the Fig. 3. (left panel), in a strongly coupled system where the Coulomb blockade peaks are suppressed, $ V_{th} $ takes on small values. This agrees with an almost featureless character of DOS shown in the Fig. 1. However, as the coupling between the QD/molecule and the electrodes weakens, the Coulomb blockade peaks appear. This brings a significant increase of thermovoltage values, and makes nonmonotonic character of $ V_{th} (\Delta\theta) $ more distinct.  The strength of Coloumb interactions also affects the thermovoltage behavior. When the charging energy $ U $ takes on certain values, the dependence of $ V_{th} $ of $ \Delta\theta $ remains  nearly linear even when $\Delta\theta $ reaches value of the order of $ \theta_R, $ and it may lack a minimum/maximum. An example of such a curve is displayed in the right panel of the Fig. 3. At other values of  $U,\  V_{th}(\Delta\theta) $ becomes an nonmonotonic function. The curves shown in the figure are plotted assuming that $ \tilde E_0 < 0. $ Accordingly, $ V_{th}(\Delta\theta) $ reveals a minimum. The minimum position depends on the value of charging energy.
  The thermovoltage behavior similar to that presented in the  right panel of Fig. 3 occurs when one considers $ V_{th}(\Delta \theta) $ at different positions of the dot energy level provided that the charging energy is fixed \cite{35}. As discussed in that work, this behavior originates from nonmonotonic dependencies of the occupation numbers on the dot of the renormalized energy $ \tilde E_0. $ Actually, $ \big<n_\sigma\big> $ is determined by the relationship between $ \tilde E_0 $ and $ U. $ Therefore, one may observe similar features in $ V_{th}(\Delta \theta) $ behavior either by varying $ E_ 0 $ at a fixed $ U $ or by varying $ U $ and keeping $ E_0 $ fixed as in the present work. 

\begin{figure}[t] %%% fig. 4
\begin{center}
\includegraphics[width=8.8cm,height=7.5cm]{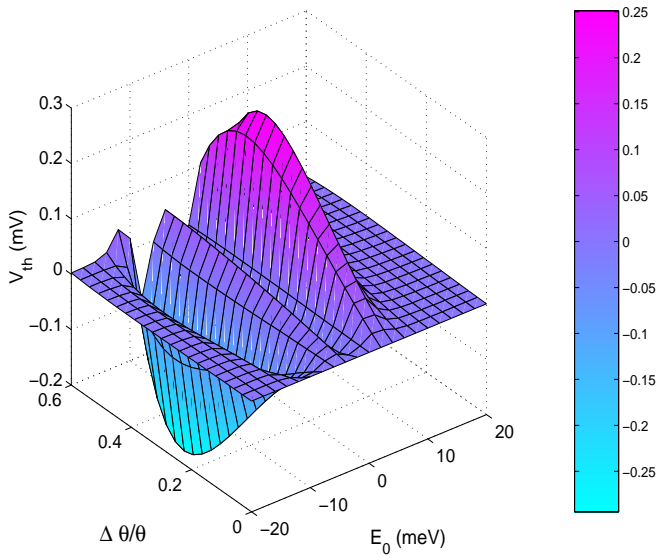} 
\caption{(Color online) The thermovoltage $V_{th} $ versus $ E_0 $ and temperature differential  $ \Delta \theta $ plotted for a weakly coupled QD $(\Gamma = 0.3meV,\ U = 6meV ) $ above the Kondo temperature $(k\theta_R = 0.6meV).$
}
 \label{rateI}
\end{center}\end{figure}

It follows from the previous analysis, that the thermovoltage, should strongly depend of the position of the dot energy level. As shown in Figs. 4-6, $ V_{th} $ rather sharply varies over the interval whose endpoints are located near $ E_0 = - U $ and $ E_0 = 0, $ respectively. Considering a system within the Kondo regime, one sees that as the dot level approaches the chemical potentials of electrodes from above, $ V_{th} $ reveals a sharp double peak shown in the left panel of the Fig. 5. Presumably, this feature appears due to the thermally excited hole transport through the system which starts when the renormalized dot energy level moves sufficiently close to the electrodes chemical potentials. The height of the peak increases as the temperature differential $ \Delta\theta $ enhances. The shape of this feature may be treated as that of a single peak with a sharp dip imbedded in the middle of the latter. The thermovoltage takes on values close to zero on the dip bottom. This fall of $ V_{th} $ probably occurs due to the effect of the Kondo maximum in the electron DOS. The presence of the latter gives rise to electron flux which counterbalances the hole flux within the corresponding range of $ E_0 $ values.
  Another and less prominent feature appears when the renormalized energy on the dot $ \tilde E_0 $ becomes close to $ -U . $ It indicates the thermally excited electron transport through the channel $ E= \tilde E_0 + U. $ 
  
   Within the Coulomb blockade regime at low temperatures the thermovoltage displays sharp features at $ \tilde E_0 = - U $ and $ \tilde E_0 = 0 $ associated with two possible channels for thermally excited transport.  Both features are characterized with derivative-like lineshapes. For the considered model representing the QD/molecule by a single orbital, the thermovoltage changes sign four times over the interval $ -U < \tilde E
_0 < 0, $ each change indicating the change of the kind of charge carriers (electrons/holes) predominating in the transport.  At higher temperatures sharp derivative-like features appearing in the $ V_{th}(E_0) $ lineshape become smoother and broader, as shown in the right panel of the Fig. 5. Nevertheless they retain essential characteristics  so long as the coupling of the electrodes to the linking QD/molecule remains weak. For stronger coupled systems the electron DOS becomes featureless, and the peaks in the $ V_{th} $ disappear, as shown in the Fig. 6.

\begin{figure}[t] %%% fig. 5
\begin{center}
\includegraphics[width=8.8cm,height=4.5cm]{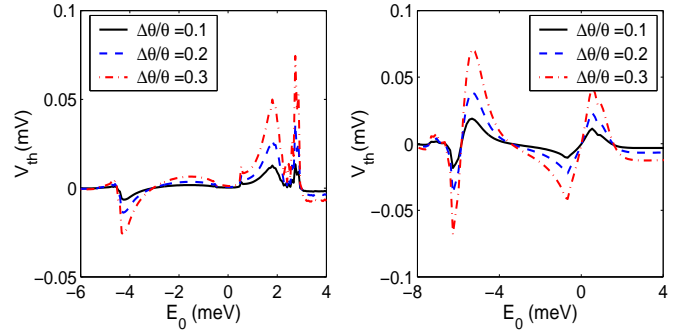} 
\caption{(Color online) Themovoltage versus $E_0 $ below the Kondo temperature. Curves are plotted assuming $ k\theta_R= 0.16meV,\ U = 6meV $ for $\Gamma = 3meV $ (left panel) and $\Gamma = 0.3meV $ (right panel).
  }
 \label{rateI}
\end{center}\end{figure}

\begin{figure}[t] %%% fig. 6
\begin{center}
\includegraphics[width=8.8cm,height=4.5cm]{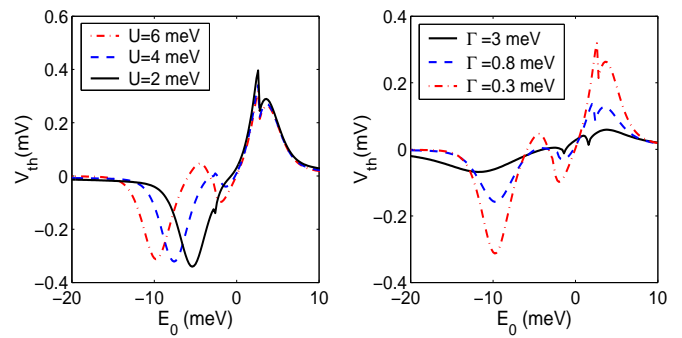} 
\caption{(Color online) Thermovoltage versus $ E_0 $ above the Kondo temperature. Curves are plotted assuming $k\theta_R = 0.6meV,\ k\theta_L=0.84meV;\ \Gamma = 0.3meV$  (left panel) and  $ U = 6meV $ (right panel).
}
 \label{rateI}
\end{center}\end{figure}

\section{iv. thermocurrent}

As mentioned above, properties of thermovoltage in nanoscale systems were studied in numerous works. Less attention was paid to studies of thermocurrent in spite of the fact that $ I_{th} $ is more convenient for measuring and modeling \cite{25}. In this Section, we present some results of theoretical analysis of the thermocurrent behavior in a system consisting of a couple of electrodes linked by a QD/molecule which is simulated by a single energy level. As before, we assume that the linker is symmetrically coupled to the electrodes, and the right electrode is cooler than the left one. The thermocurrent is defined as follows:
\be
 I_{th} = I(V,\theta_R, \Delta\theta) - I (V,\theta_R, \Delta\theta = 0)                          \label{14}
\ee
where $ V $ is the bias voltage  applied across the system. According to this definition, $ I_{th} $ represents the contribution to the tunnel current which originates from thermally excited transport of charge carriers. Employing Eq. (\ref{12}), $ I_{th} $ may be presented in the form:
\begin{align}
I_{th} = & \frac{e}{\pi\hbar} \int \Big\{ T(E,\theta_R,\Delta\theta) f^L(E,\theta_L)
\nn\\   &-
T   (E,\theta_R,\Delta\theta =0)f^L(E,\theta_L) - \Delta T f^R(E,\theta_R)\Big\} dE.                                                  \label{15}
\end{align}
Here, $\Delta T $ is the difference between the electron transmission function computed at nonzero value of the temperature differential $ \Delta \theta $ and that one computed assuming $ \Delta\theta = 0: $
\be
  \Delta T =  T(E,\theta_R,\Delta\theta) - 
T   (E,\theta_R,\Delta\theta =0)        .                     \label{16}
\ee
In these expressions (\ref{15}), (\ref{16}), the electron transmission is given by Eq. (\ref{13}).

\begin{figure}[t] %%% fig. 7
\begin{center}
\includegraphics[width=8.8cm,height=7.5cm]{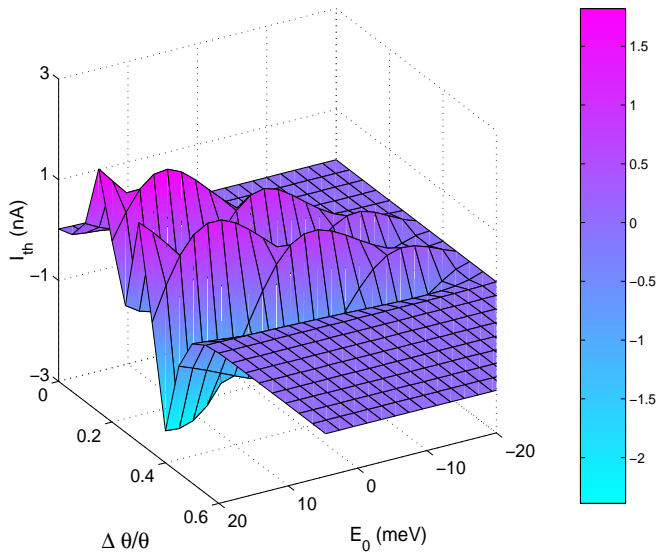} 
\caption{(Color online) Thermocurrent $ I_{th} $ flowing through an unbiased QD as a function of the energy $ E_0 $ and temperature differential $ \Delta\theta .$ The surface is plotted assuming $ \Gamma = 0.3 meV,\ k\theta_R = 0.6meV,\ U = 6meV. $
}
 \label{rateI}
\end{center}\end{figure}

The value of $ I_{th} $ is determined by the magnitude and polarity of the bias voltage $V,$  and by the temperature differential $ \Delta\theta. $ Other factors strongly affecting $ I_{th} $ include the position of the energy level on the QD/molecule $ E_0, $ the coupling parameter $ \Gamma,$ the value of charging energy $ U $ and the average temperature $ \theta. $ We first consider the case of an unbiased system$(V = 0).$ In this case, $ I_{th }$ solely originates from the thermally excited flow of charge carriers. The thermocurrent dependence of   $ E_0 $ and $\Delta\theta $ is shown in the Fig. 7. The flow occurs when the energy level on the dot is  shifted to a position where the renormalized energy $ \tilde E_0 $ (or $\tilde E_0 + U )$ is close to the chemical potential of electrodes. The intensity of the flow determines $ I_{th} $ value. It increases as the temperature differential $ \Delta \theta $ enhances. This is illustrated in the Fig. 8 where the behavior of $ I_{th} $ as a function of $ E_0 $ is shown for several values of $ \Delta \theta. $ The results presented in this figure are corresponding to a weakly coupled system $ (\Gamma \ll U) $ considered at low $(\theta < \theta_k) $ and moderately high $(\theta > \theta_k) $ temperatures.

\begin{figure}[t] %%% fig. 8
\begin{center}
\includegraphics[width=8.8cm,height=4.5cm]{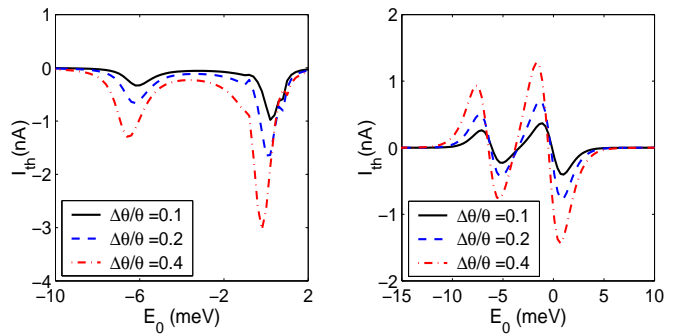} 
\caption{(Color online) Thermocurrent versus $ E_0 $ for an unbiased QD within the Coulomb blockade regime below (left panel; $ k\theta_R = 0.16meV) $ and above (right panel;  $ k\theta_R = 0.6meV) $ Kondo temperature. The curves are plotted assuming $ \Gamma = 0.3meV,\ U = 6meV. $
}
 \label{rateI}
\end{center}\end{figure}

At low temperatures, $I_{th} $ takes on zero value everywhere except close vicinities of two points indicating the opening of transport channels for the thermally excited transport. In these vicinities, the thermocurrent behavior is characterized by sharp and narrow dips whose heights increase as $ \Delta\theta $ enhances. Negative values of $ I_{th} $ show that the charge carriers involved in the transport process are electrons. At higher temperatures,  $ I_{th} $ behavior is somewhat modified. As before, the thermocurrent takes on nonzero values near $ \tilde E_0 = - U $ and $ \tilde E_0 = 0 $ which correspond to two transport channels typical for a system with a single-state bridge within the Coulomb blockade regime. However, the relevant features now have derivative-like lineshapes. So, in this case charge carriers of both kinds participate in transport via the transport channels. At certain values of $ E_0$ the two  fluxes counterbalance each other, so $ I_{th} $ becomes zero. Again, the enhancement of temperature differential results in the increase of the features heights.

The dependencies of the thermally excited current flowing through an unbiased QD/molecule of the temperature differential are strongly influenced by the position of the energy level $ E_0 .$  This is illustrated by the curves displayed in the left panel of the Fig. 9. These curves are plotted for a system within the Coulomb blockade regime at a moderately high temperature $ (\theta > \theta_k).$ Thermally excited transport in such systems was recently studied \cite{35}, and the reported results for $ I_{th} $ qualitatively agree with those presented here. As shown in the Fig. 9,  $ I_{th} (\Delta\theta) $ reaches its minimum at a certain value of $ \Delta\theta $ for each considered value of $ E_0 .$  Besides, some of these functions show maxima at small values of $ \Delta \theta $ with the subsequent change of sign. As well as in the previously discussed case of thermovoltage, the variety of $ I_{th}(\Delta\theta) $ lineshapes originates from the specific nonmonotonic dependencies of the occupation numbers on the QD/molecule energy level.

\begin{figure}[t] %%% fig. 9
\begin{center}
\includegraphics[width=8.8cm,height=4.5cm]{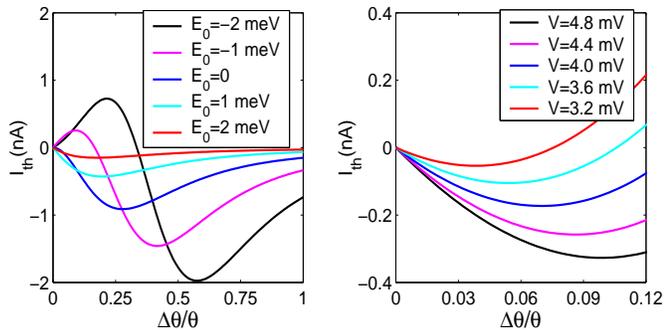} 
\caption{(Color online) Thermocurrent as a function of $ \Delta\theta $ within the Colomb blockade regime above Kondo temperature. The curves are plotted assuming $ k\theta_R = 0.6 meV,\  \Gamma=0.3meV,\ U = 6meV $ for an unbiased (left panel) and biased (right panel) quantum dot. 
}
 \label{rateI}
\end{center}\end{figure}

When a bias voltage is applied across the system, the thermocurrent is simultaneously driven by this voltage and the temperature differential. The combined effect of these two factors determines the $ I_{th} $ behavior. In general, $ \Delta \theta $ remains predominating at small values of the bias voltage $ V. $ However, for stronger  bias, its effect strengthens to such extent that $ I_{th} $ approaches zero regardless of $ \Delta\theta $ value. In a biased system, the chemical potentials of the electrodes are  shifted away from previously occupied positions,  and the conduction window opens up. Assuming, as before, that in the unbiased system chemical potentials associated with both electrodes equal zero, we may present them in the form: $ \mu_L = \alpha|e|V;\ \mu_R = -(1- \alpha)|e|V.$ In these expressions, $ \alpha $ is the division coefficient which shows  how the bias voltage is distributed between the electrodes. In further analysis we put $ \alpha = 1/2 $ which seems natural for a symmetrically coupled system.

\begin{figure}[t] %%% fig. 10
\begin{center}
\includegraphics[width=8.8cm,height=4.5cm]{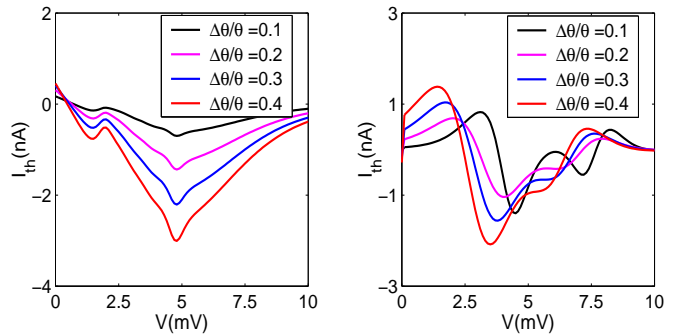} 
\caption{(Color online) Thermocurrent as a function of the bias voltage $ V $ in a single-level quantum dot within the Kondo regime (left panel) and within the Coulomb blockade regime (right panel). Curves are plotted assuming $E_0 = -2meV,\ U = 6meV $ for $  \Gamma = 3meV,\ k\theta_R  = 0.16meV $   (left panel) and for   $\Gamma =0. 3meV,\ k\theta_R  = 0.6meV $ (right panel).
}
 \label{rateI}
\end{center}\end{figure}

When the conduction window sufficiently broadens to contain transport channels, charge carriers start to tunnel through the system. However, this transport is affected by the influence of the temperature differential. Due to the presence of the latter, two terms in the expression (\ref{14}) do not cancel  each other, and $ I_{th} $ may take on nonzero values. Curves shown in the right panel of Fig. 9 are plotted for several close values of the bias voltage. These values are chosen in such a way that the energy level on the dot is situated near the boundary of the conduction window, either within or barely outside it. Therefore, the transport channel at $ E = \tilde E_0 $ appears near the shifted chemical potential of the hot electrode. This particular position of the dot energy level provides the most favorable opportunities for the temperature differential $\Delta\theta $ to influence the transport by broadening the Fermi distribution function associated with the hot electrode. Besides, temperature differential affects the transport by renormalizing the energy $ E_0. $ As a result, all presented curves display nonlinear lineshapes resembling those observed in recent experiments on semiconducting quantum dots \cite{25}, and these nonlinear properties of thermoelectric transport appear even at small values of $  \Delta \theta. $

Some further characteristic properties of thermocurrent flowing through biased QD/molecules are illustrated in the Fig. 10. In this figure, we show $ I_{th} $ as a function of the bias voltage for a system within the Kondo regime (left panel) as well as within the Coulomb blockade regime (right panel). As known, the bias voltage application leads to a splitting of the Kondo peak in two maxima which are gradually suppressed as the bias increases. The curves displayed in the left panel of the Fig. 10 are plotted assuming that bias voltage takes on  low values, and the Kondo peak, although affected, is not completely dissolved. Accordingly,  each $I_{th} -V $ curve presented in the left panel is characterized by two minima. These features appear due to the presence of two peaks in the electron DOS replacing a single Kondo maximum in a slightly biased QD. 

When the coupling between the electrodes and the linking QD/molecule weakens, and two peaks associated with the Coulomb blockade regime emerge, the lineshapes of $I_{th} - V $ characteristics significantly change. The interplay between electric and thermal driving forces brings several $ I_{th} $ sign reversals, as shown in the right panel of Fig. 10. One may suggest that these reversals and accompanying derivative-like features occur as a result of opening up transport channels at $ E = \tilde E_0,\ \tilde E_0 + U $ within the conduction widow. Comparing Figs. 9 and 10, one may observe similarities in  $I_{th} $ behavior in unbiased and slightly biased systems within the Coulomb blockade regime. In both cases derivative like features appear in the $ I_{th} $ lineshapes indicating the opening of transport channels, although these features shapes in biased systems are distorted due to the bias voltage influence. When the bias further increases, electric driving forces become predominating. Then the difference between the terms in Eq. (\ref{14}) diminishes and $ I_{th} $ approaches zero as illustrated in Fig. 10.

\section{v. conclusion}

Finally, we repeat again that thermoelectric properties of nanoscale systems attract significant interest of the research community. The present work was inspired by this common interest as well as by recent observations of strongly nonlinear Seebeck effect in semiconducting quantum dots  bridging two electrodes \cite{25}. Coulomb repulsion between electrons on the bridge may significantly affect characteristics of thermoelectric transport through such systems. The effects of electron-electron interactions were theoretically analyzed in several works (See e.g. Refs. \cite{15,22,26,27,35}. However, in these works the analysis was carried out within the Coulomb blockade regime $(\Gamma\ll U).$ Here, we employ a computational scheme within the EOM approach which enables to study thermoelectric transport through QD/molecules sufficiently strongly coupled to electrodes.

We realize that EOM approach used in this work to compute relevant Green's functions has its limitations. In principle, better approximations for Green's functions appropriate for studies of electrical and thermal transport in Kondo correlated systems could be derived by applying a numerical renormalization group approach. Nevertheless, as discussed in an earlier work \cite{20}, EOM based approximations for the Green's functions used in the present work bring reasonably good results for transport characteristics assuming that the considered system is not too strongly coupled, so that $ \Gamma $ remains smaller that $ U. $ 

Applying the suggested computational scheme, we showed that the thermovoltage created by temperature differential may decrease in magnitude when $ \Delta\theta $ increases. At certain values of $ \Delta\theta,\ V_{th} $ even may become zero and  change its polarity as $ \Delta\theta $ increases beyond these values. This interesting phenomenon was observed in experiments and theoretically analyzed for a case of a QD within the Coulomb blockade regime \cite{35}.  Here, we showed that $ V_{th} $ may similarly behave in a strongly coupled system kept below Kondo temperature.
However, this behavior of thermovoltage is universal in character. In general, $V_{th} $ behavior  is determined by interplay of several factors including the position of the energy level on the considered single-level dot, the average temperature, the coupling parameter $ \Gamma $ and the charging energy $ U. $ We have analyzed some effects of these factors. For instance,  we showed that at certain values of relevant energies, $ V_{th} (\Delta\theta) $ becomes a monotonic function which agrees with results obtained in several earlier works (See e.g.  Refs. \cite{22,27}).

Another important characteristic of thermoelectric transport is thermally excited electric current. This quantity is especially interesting for it is available for direct measuring in experiments on nanoscale systems. The thermoelectric current $ I_{th} $ flowing through an  unbiased QD/molecule originates from the flux of charge carriers from the hot electrode to the cooler one. Within the accepted model where the dot/molecule is represented by a single orbital, there exist two channels for the thermally excited tunnel transport. These channels open when either the energy on the dot $ E_0 $ or $E_0 +U $ approaches the electrodes chemical potentials. Therefore, the value of the charging energy $ U $ is an important factor affecting $ I_{th} $ behavior.  The effect of Coulomb interactions strongly depends on the relationship between the energies $ U $ and $ \Gamma $ and on the average thermal energy $ k\theta. $ Varying these energies, we can move the considered system from Kondo regime to Coulomb blockade regime and study specific manifestations of Coulomb interactions within these regimes. The corresponding analysis is carried out in the present work. In a biased QD/molecule, the thermocurrent behavior is determined by the interplay of the bias voltage and thermal driving forces.  Coulomb interactions affect $ I_{th} $ behavior in biased systems, as well. We have analyzed the effect of Coulomb interactions on the thermocurrent flowing through a biased QD within the accepted model for the latter.  We believe that presented computational method and obtained results may be helpful for further studies of thermoelectric properties of nanoscale systems.

\vspace{2mm}

 {\bf Acknowledgments:}
The author  thank  G. M. Zimbovsky for help with the manuscript. This work was partly supported  by  NSF-DMR-PREM 0353730.

\end{document}